\documentclass[journal=jpclcd,manuscript=letter,articletitle=true,layout=twocolumn]{achemso}
\usepackage{amsmath}
\usepackage{amssymb}
\usepackage{bm}
\usepackage{latexsym}
\usepackage{graphicx}
\usepackage{epsfig,epsf,rotating}
\usepackage{indentfirst}
\usepackage{epstopdf}
\usepackage{xcolor}
\usepackage{multirow}
\usepackage{tabularx}
\usepackage{ctable}

\usepackage{upgreek}
\newcolumntype{C}{>{\centering\arraybackslash}X}
\usepackage{nicefrac}

\title{Equilibration between Translational and Rotational Modes in Molecular Dynamics Simulations of Rigid Water Requires a Smaller Integration Time-Step Than Often Used\footnote{
Notice:  This manuscript has been authored by UT-Battelle, LLC, under contract DE-AC05-00OR22725 with the US Department of Energy (DOE). The US government retains and the publisher, by accepting the article for publication, acknowledges that the US government retains a nonexclusive, paid-up, irrevocable, worldwide license to publish or reproduce the published form of this manuscript, or allow others to do so, for US government purposes. DOE will provide public access to these results of federally sponsored research in accordance with the DOE Public Access Plan (https://www.energy.gov/doe-public-access-plan}}
\author{Dilipkumar N. Asthagiri}
\affiliation{Oak Ridge National Laboratory, One Bethel Valley Road, Oak Ridge, TN 37830-6012}
\email{asthagiridn@ornl.gov}
\author{Thomas L. Beck}
\affiliation{Oak Ridge National Laboratory, One Bethel Valley Road, Oak Ridge, TN 37830-6012}
\email{becktl@ornl.gov}

\begin{document}
\begin{abstract}
In simulations of aqueous systems it is common to freeze the bond vibration and angle bending modes in water to allow for a longer time-step $\delta t$ for integrating the equations of motion. Thus $\delta t = 2$~fs is often used in simulating rigid models of water. We simulate the SPC/E model of water using $\delta t$ from 0.5~fs to 3.0~fs. We find that for all but $\delta = 0.5$~fs, equipartition between translational and rotational modes is violated: the rotational modes are at a lower temperature than the translation modes. The autocorrelation of the velocities corresponding to the respective modes shows that
the rotational relaxation occurs at a time-scale comparable to vibrational periods, invalidating the original assumption for freezing vibrations.  $\delta t$ also influences thermodynamic properties:  the mean system potential energies are not converged until $\delta t = 0.5$~fs, and the excess entropy of hydration of a soft, repulsive cavity is also sensitive to $\delta t$. 
\begin{tocentry}
\vspace{5mm}
\center{\includegraphics[scale=0.625]{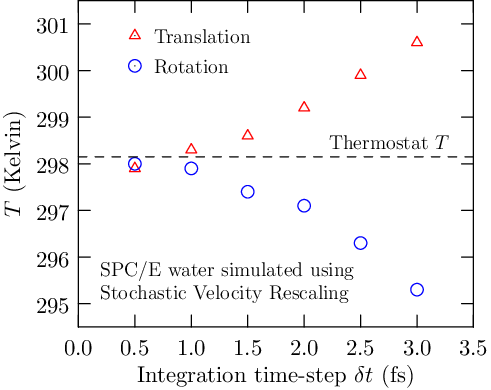}}
\end{tocentry}
\end{abstract}
 
Water is the matrix of life, and a molecular level understanding of biological processes is predicated on understanding how the bio-molecules are influenced by the liquid water matrix \cite{ball:cr2008}. Thus, developing better ways to describe the structure and dynamics of water has come to occupy a central position in computational (bio)molecular sciences.

Rahman and Stillinger, the early pioneers in simulating water using molecular dynamics, described water as a rigid object 
and numerically solved for the coupled translational and rotational motion of each water molecule in the liquid\cite{rahman:jcp71,stillinger:jcp74}. In their numerical scheme, the translational motion of the center of mass was formulated in terms of Cartesian coordinates and the rotational motion was formulated in terms of Euler angles. The Hamiltonian of the system was based on a sum of Lennard-Jones and electrostatic contributions. Using the mass $m$ of the water molecule, the Lennard-Jones well-depth $\varepsilon$, and collision diameter $\sigma$, they identified the natural unit of time in the equations of motion to be $\tau = \sigma \cdot \sqrt{m/\varepsilon} \approx 2$~ps \cite{rahman:jcp71}. From numerical experiments on a two-molecule system, they found that to integrate the coupled set of equations, they had to use a time-step $\delta t = 2\times 10^{-4} \tau \approx 0.4$~fs. They noted that the smallness of this time-step, relative to modeling liquid Ar, for example, stemmed from the ``rapid angular velocity of the water molecules" \cite{rahman:jcp71}. We shall return to this point below.  

A few years after Rahman and Stillinger's work, Ryckaert, Ciccotti, and Berendsen developed the SHAKE algorithm\cite{md:shake} to incorporate holonomic constraints in simulating various types of molecules, including water. For simulating a rigid model of water, one choice of constraints could be OH bond lengths and a pseudo-bond between the two protons. These holonomic constraints lead to additional forces in the dynamical equations, but the significant computational advantage one gains is in formulating all the equations in Cartesian coordinates. Later, Andersen noted that in simulating a rigid object,  the relative velocity of atoms mutually tied by a rigid bond should be zero in the direction of the bond. Andersen developed the RATTLE algorithm to include this velocity constraint \cite{andersen:rattle83}.  More than a decade after Andersen's work, Miyamoto and Kollman \cite{kollman:settle92} presented an algorithm specifically for describing water as a rigid molecule. This so-called SETTLE algorithm obviated the iterations implicit in the SHAKE or RATTLE methods.  Like SHAKE, the RATTLE and SETTLE methods require only Cartesian coordinates. 

One of the motivations in developing the SHAKE and subsequent algorithms noted above was to freeze the high-frequency vibrations between bonded pairs of atoms, since the ``fast internal vibrations are usually decoupled from rotational and translational motions'' \cite{md:shake}. Since typical vibrational frequencies are about $10^{14}$ Hz (or a time period of 10~fs), the intuitive idea was that freezing the high-frequency vibrations ought to allow for a longer $\delta t$ for integrating the equations of motion in molecular simulations. Thus, in the case of simulating rigid water using molecular dynamics, it is very common to use $\delta t = 2$~fs; for example, see Refs.\ \citenum{Hummer:nature01,paschek:jcp04,garde:pnas09}. We have used this as well in many of our papers, for example, Ref.\ \citenum{tomar:jpcl20}. A well-cited paper on protein folding 
has used $\delta t = 2.5$~fs \cite{deshaw:sc2011}. Some recent efforts use $\delta t = 4$~fs, albeit by using a larger proton mass (personal communications, CECAM 2023 meeting on Biomolecular Simulation and Machine Learning in the Exa-Scale era). 

What is the problem then? In simulation studies on liquid water under super-cooled conditions using the Langevin thermostat (Valiya Parambathu and Asthagiri, unpublished), 
 one of us (DNA) noticed that the average temperature from the simulation log was systematically lower than the target by about 1~K. We found the same trend in simulations of hydration of a small amphiphile, {\it tert}-butanol. This motivated us to examine more thoroughly the distribution of kinetic energy between translational and rotational degrees in simulations of the rigid SPC/E\cite{spce} water model. This examination leads to the finding that for $\delta t \geq 1$~fs, equipartition between translation and rotation is violated, with the problem made worse as $\delta t$ increases.  We examine the reason for this and find that $\delta t \geq 1$ limits how well one can capture the fast rotational  relaxation. The  $\delta t = 0.5$~fs that allows thermalization between rotation and translation is also surprisingly close to what  Rahman and Stillinger\cite{rahman:jcp71} used. The violation of equipartition also reveals itself in the $\delta t$ dependence of the mean potential energy of the system and in the excess entropy of hydration of a soft, repulsive cavity.  

We study the SPC/E\cite{spce} water model using the NAMD\cite{namd,namd:2020} and LAMMPS\cite{plimpton:jcop1995,lammps:2022} codes. For NVT simulations, we obtain results using the stochastic velocity rescaling \cite{svr:jcp07}  and Langevin thermostats, respectively, both of which should correctly sample the canonical distribution. We study two system sizes, $N=2006$ water molecules, which informs the results in the main text, and a limited set of simulations with $N=16384$ water molecules noted in the Supplementary Information (SI). The SI provides a complete description of the methods (SI Sec.\ 2) and additional supporting results (SI Sec.\ 3). 

From $NVT$ simulations we sample 1050 water molecules (SI Sec.\ 2.2). From the coordinates and velocities of the atoms at a given time point for a sampled water molecule,  we calculate the translational kinetic energy of the center of mass and the kinetic energy for rotation about the center of mass.  For a canonical ensemble, these kinetic energy contributions must follow the Maxwell-Boltzmann distribution: specifically, 
for a single degree of freedom, the mean kinetic energy is $k_{\rm B}T / 2$ and the variance is $(k_{\rm B}T)^2/2$. This allows us to calculate $T$ following two different paths (SI Sec.\ 1). The different paths lead to the same estimate of $T$ within statistical uncertainties, as they should for well-converged simulations.  

\begin{figure*}[h!]
\includegraphics[scale=0.95]{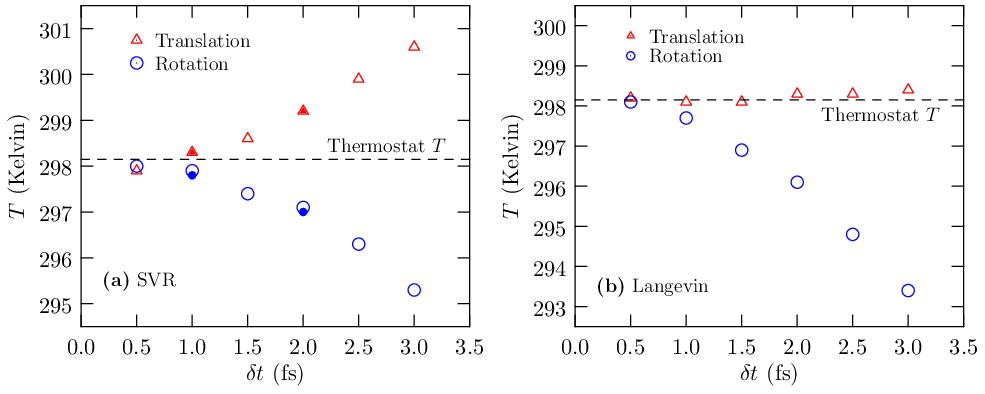}
\caption{Temperature distribution between the translational and rotational modes in simulations of SPC/E water. The temperatures are based on the mean kinetic energy path.  Results using (a) SVR and (b) Langevin thermostats. The open symbols are based on simulations with NAMD \cite{namd, namd:2020}. The symbol size is about $2\sigma$ standard error of the mean. The filled symbols in (a) are based on simulations using LAMMPS \cite{plimpton:jcop1995,lammps:2022}. } 
\label{fg:TransRotK}
\end{figure*}

Figure~\ref{fg:TransRotK} shows that the temperatures ascribable to the different modes converge only for $\delta t = 0.5$~fs. 
For the Langevin thermostat with $\delta t = 2.0$~fs, the mean of the translational and rotational temperatures is about 297~K, a Kelvin lower than the set-point, consistent with the earlier observations for supercooled water and the hydration of the amphiphile that motivated the present work. Similar deviations persist in constant energy (NVE) simulations as well (SI Sec.\ 3.4), emphasizing that the data in Figure~\ref{fg:TransRotK} is not an artifact of the thermostat. 

\begin{figure*}[h!]
\includegraphics[scale=0.9]{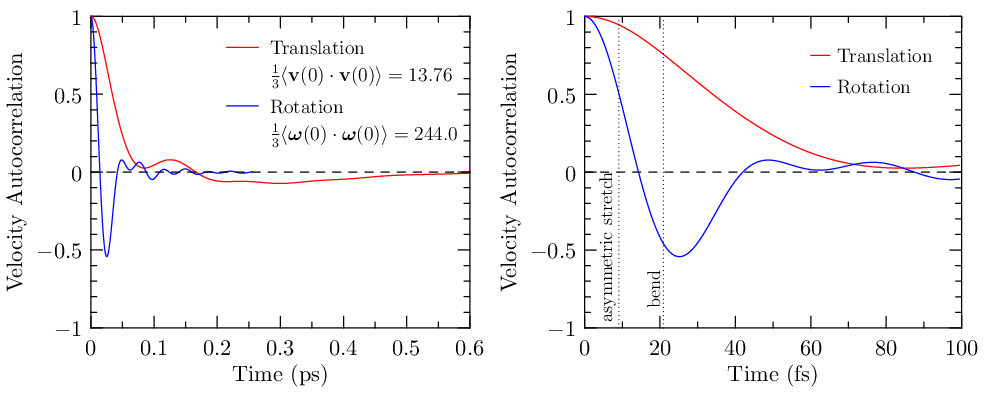}
\caption{Autocorrelation of the center of mass and angular velocities normalized by the value at time $t = 0$. Left panel: data shown out to 0.6~ps. Right panel: data shown to 100~fs. The time periods associated with the asymmetric stretch and bending modes of a water molecule in the gas phase \cite{kananenka:jcp18} are also shown for comparison.} 
\label{fg:ACF}
\end{figure*}

To better understand the results shown in Figure~\ref{fg:TransRotK}, it proves helpful to examine the velocity autocorrelation function. To this end, we take the last configuration of the $\delta t = 0.5$~fs run with the SVR thermostat and launch constant energy runs for 20~ps saving velocities and positions every time step. For NVE starting velocities, we use the same seed for the random number generator for the different $\delta t$ cases. Figure~\ref{fg:ACF} shows the velocity autocorrelation using the data from the $\delta t = 0.5$~fs simulations, our reference. In passing, we note that the integral of the velocity autocorrelation gives the diffusion coefficient through the Green-Kubo relations. As a check, for $\delta t = 0.5$~fs we find the translational diffusion coefficient using both the Green-Kubo approach and the Einstein relation (SI Sec.\ 3.6) for the mean squared displacement. These estimates agree within statistical uncertainties. 

What is striking about Fig.~\ref{fg:ACF} (left panel) is that the rotational motion relaxes considerably faster than the translational motion --- observe that the $t = 0$ value is much higher and the decay much faster for the autocorrelation of the angular velocity. This is consistent with what Rahman and Stillinger\cite{rahman:jcp71} noted.  Physically, the rotational relaxation is considerably faster than the translation relaxation because, in contrast to the 
translational movement of the entire mass, the small mass of the proton relative to that of the oxygen means that the moment of inertia is small and rotations are sensitive to small torques. 

As Fig.~\ref{fg:ACF} (right panel) shows, the rotational relaxation occurs over time scales that are similar to the time period of the bond vibration and angle bending modes of a water molecule in the gas phase. For a water molecule in the liquid, because of hydrogen bonding interactions these modes will be ``softened" and shift to the right and spread out \cite{kananenka:jcp18}. So, even using  the worst case estimate of a water molecule in the gas phase, we find that the original motivation for freezing vibrations is not well-founded for describing liquid water. 

\begin{figure*}[h!]
\includegraphics[scale=0.95]{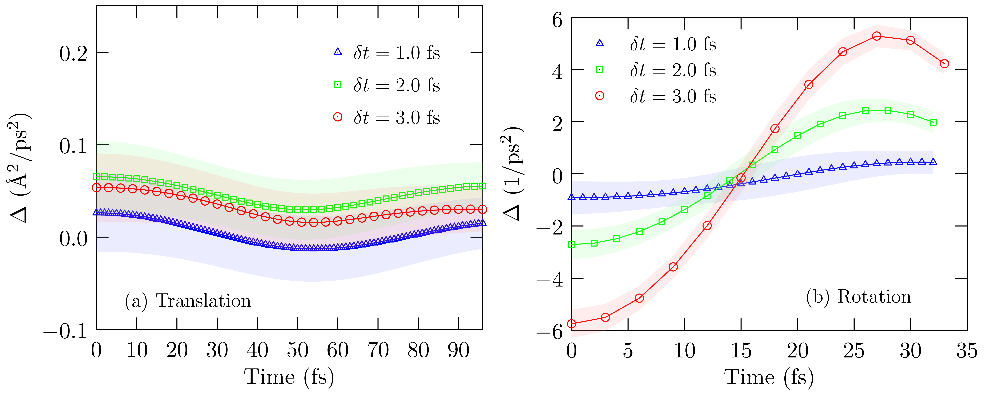}
\caption{$\Delta$ is the difference at a given time point between the autocorrelation for the given $\delta t$ minus the corresponding reference value using $\delta t = 0.5$~fs. For no discretization error, $\Delta = 0$. (a): Translational, and (b): Rotational motion. The shaded area indicates $1\sigma$ standard error of the mean.} 
\label{fg:relACF}
\end{figure*}
Figure~\ref{fg:relACF} shows the difference between the velocity
autocorrelation  obtained using a given $\delta t$ and the reference value based on $\delta t = 0.5$~fs (Fig.~\ref{fg:ACF}). We focus on the initial times as this is the more important and more sensitive part of the overall relaxation. 
It is immediately clear that discretization of the equations of motion limits the fidelity with which we can assess the corresponding relaxation.  For the translational motion (Fig.~\ref{fg:relACF}, left panel), the differences at initial times are all positive. 
However, the net impact on the diffusion coefficient as assessed by the Einstein relation is small (SI Sec.\ 3.6); this suggests that integrating the overall autocorrelation can smooth out errors and mask the role of $\delta t$.  Similarly, for the rotational motion (Fig.~\ref{fg:relACF}, right panel), we find that the autocorrelation up to about 15~fs is lower  than the reference value.  
Notice the stark contrast in the magnitude of deviations relative to the reference autocorrelation. Since the rotational relaxation occurs on a considerably shorter time scale (and involves larger magnitudes), it is also more sensitive to discretization: large $\delta t$ leads to large errors in describing rotational relaxation than it does for translation relaxation. 

We can use the fluctuation-dissipation relation $\mathcal{D} = k_{\rm B}T / \xi$, where $\mathcal{D}$ is the diffusion coefficient, $\xi$ is the friction, and $k_{\rm B}T$, to interpret the results above. Treating $\xi / k_{\rm B}T$  as an ``effective" friction, and solely focusing on the initial time behavior of the autocorrelation, we can infer that the ``effective" friction appears to be higher for rotational motion and lower for translational motion relative to the $\delta t = 0.5$ fs reference. Friction arises due to inter-molecular interactions. Since the same potential model is used for all $\delta t$ values, a high ``effective" friction is a consequence of lower temperature and vice versa, in agreement with results in Fig.~\ref{fg:TransRotK}. 

We hypothesized that the discrepancy in thermalization between rotational and translational motions ought to influence properties besides initial relaxation, including thermodynamic properties such as the excess entropy. To test this hypothesis, we first 
calculated the mean binding energy, $\langle \varepsilon\rangle$, of a water molecule with the rest of the system (SI Sec.\ 2.1.4) --- the mean potential energy of the $N$ water molecule system is $N \langle \varepsilon\rangle / 2$. Fig,~\ref{fg:TransRotK} (left panel) shows that the mean binding energy is indeed sensitive to $\delta t$, with the values obtained using different thermostats converging only for $\delta t = 0.5$~fs. Specifically, for $\delta t = 3$~fs, the mean potential energy \underline{per water molecule} differs by 0.3\% between the two thermostats; for $\delta t = 0.5$~fs the deviation drops to 0.03\%. 
\begin{figure*}
\includegraphics[scale=0.95]{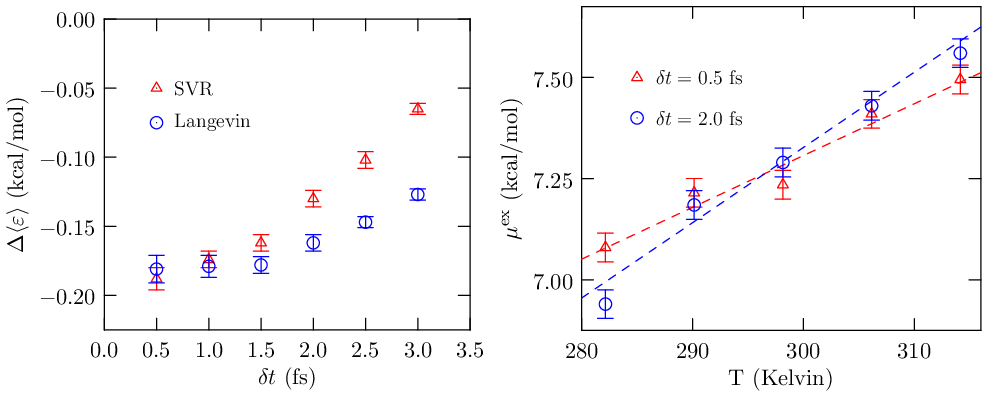}
\caption{\underline{Left panel}: Mean binding (interaction) energy of a water molecule with the rest of the system. $\Delta \langle \varepsilon \rangle = \langle \varepsilon \rangle - 22$.  The standard error of the mean value is at the 2$\sigma$ level. Note that a deviation of 0.1~kcal/mol per particle amounts to $\approx 200$~kcal/mol for the entire system. \underline{Right panel}: $\delta t$ dependence of the hydration free energy of a soft, repulsive cavity.  The standard error of the mean is shown at 1$\sigma$ level.} 
\label{fg:PE}
\end{figure*}
Fig.~\ref{fg:PE} (right panel) shows the hydration free energy of a soft, repulsive cavity of size 4~{\AA}  (SI Sec.\ 2.3) as a function of temperature. The excess entropy of hydration at 298.15~K obtained from a linear fit to the $\mu^{{\rm ex}}$ data is $-12.8\pm 0.1$~cal/mol-K for $\delta t = 0.5$~fs and $-18.6\pm 0.1$~cal/mol-K for $\delta t = 2.0$~fs, a nearly 50\% difference. The free energies themselves are about the same at 298.15~K, suggesting compensation between enthalpic and entropic effects. 

The identification of fast rotational relaxation of water by us is certainly not new. For example, an early pioneering study by Lawrence and Skinner\cite{skinner:jcp2002a} (see their Figure 4) conveys this point, as does Figure 2 in this work. 
 It is encouraging, and also humbling, that Rahman and Stillinger\cite{rahman:jcp71} chose a small $\delta t$ to better describe the fast angular relaxation, especially at a time when simulations were a lot more demanding than they are now.  However, the key finding of our work is that not capturing the fast rotational relaxation of water can break equipartition itself. Further, our work shows that the relaxation of the bond vibrations of water occur at a time scale similar to the rotational relaxation in the rigid model of water studied here, and thus the original motivation for using SHAKE (or SETTLE) for water can be questioned. 

Earlier researchers\cite{toxvaerd:pre94,gans:pre2000,shaw:jctc2010} have shown that in discrete Hamiltonian dynamics the evolution of the system follows a so-called shadow Hamiltonian, which is a function of $\delta t$ (SI Sec.\ 3.3). This  confounds the relationship between momentum (which is what one calculates using the ``velocity"-Verlet equations) and velocity in classical mechanics\cite{gans:pre2000}. Thus it has been argued that the Velocity Verlet velocities are not the most appropriate quantities for estimating the temperature \cite{gans:pre2000,shaw:jctc2010}. Despite this, the on-step ``Velocity" Verlet velocities are the quantities used in most codes for calculating the instantaneous temperature. Importantly, our identification of the breakdown of equipartition (Fig.~\ref{fg:TransRotK}), and the thermodynamic consequences of using a large step size (Fig.~\ref{fg:PE})  does emphasize the need to use a smaller $\delta t$ than has been used in numerous studies of liquid water. Our study here also suggests caution in using larger step sizes in aqueous bio-molecular simulations by changing the proton mass (SI Sec.\ 3.5). 

We leave it to future studies to re-examine earlier significant results in aqueous phase chemistry and biology that are founded on molecular simulations using a large time step. The findings here are also expected to be relevant to the modeling of rate processes that are sensitive to solvent friction, to efforts in coarse-graining simulations, and in the development and benchmarking of forcefields.

\section{Supporting Information}

Supporting Information includes (1) a brief discussion of how temperature is related to the kinetic energy and the fluctuation of kinetic energy, (2) methodology --- simulation systems, codes, time steps, calculation of translational and rotational kinetic energy, and calculation of the hydration free energy of a soft, repulsive cavity, and (3) Supplemental results --- tabulated data for results in Figure 1,  data for relative velocity along constrained bonds, sampling of the shadow Hamiltonian, translational and rotational energy in NVE simulations, influence of the proton mass for $\delta t = 4$~fs, and mean-squared displacement for translational diffusion coefficient. 

\section{Acknowledgements}
We thank Arjun Valiya Parambathu (U.\ Delaware), Thiago Pinheiro dos Santos (Rice University), Lawrence Pratt (Tulane University), Van Ngo (ORNL), and David Rogers (ORNL) for helpful discussions, and Nick Hagerty (OLCF) for help with LAMMPS on Summit and Frontier supercomputers.  This research used resources of the Oak Ridge Leadership Computing Facility at the Oak Ridge National Laboratory, which is supported by the Office of Science of the U.S. Department of Energy under Contract No. DE-AC05-00OR22725.

\providecommand{\latin}[1]{#1}
\makeatletter
\providecommand{\doi}
  {\begingroup\let\do\@makeother\dospecials
  \catcode`\{=1 \catcode`\}=2 \doi@aux}
\providecommand{\doi@aux}[1]{\endgroup\texttt{#1}}
\makeatother
\providecommand*\mcitethebibliography{\thebibliography}
\csname @ifundefined\endcsname{endmcitethebibliography}
  {\let\endmcitethebibliography\endthebibliography}{}

\end{document}


\newpage

\tableofcontents

\clearpage

\section{Temperature in classical statistical mechanics}
Consider a classical ensemble of $N$ particles at a specified volume $V$ and in equilibrium with a thermal bath at temperature $T$. The physical Hamiltonian of the system $\mathcal{H}$ comprises the kinetic energy and the potential energy. The kinetic energy of each particle is quadratic in the momenta. 

Consider for simplicity a single degree of freedom. Then the probability density distribution of kinetic energies $\varepsilon_K$ follows the well known Maxwell-Boltzmann form 
\begin{eqnarray}
f(\varepsilon_K)= \frac{1}{\sqrt{\pi k_{\rm B}T}}\cdot \frac{1}{\sqrt{\varepsilon_K}} \cdot \exp(-\frac{\varepsilon_K}{k_{\rm B}T}) \, ,
\label{eq:MB}
\end{eqnarray}
 where $k_{\rm B}$ is the Boltzmann constant and $T$ is the temperature of the thermal bath.  This probability density distribution depends on only one parameter $T$. For a given set of 
 observations of the kinetic energy, let the mean kinetic energy be $\bar{\varepsilon}_K$. 
 If these observations obey Eq.~\ref{eq:MB}, we require
  \begin{eqnarray}
 \frac{k_{\rm B} T}{2} = \bar{\varepsilon}_K \Rightarrow T = \frac{2 \bar{\varepsilon}_K}{k_{\rm B}}
 \label{eq:mean}
 \end{eqnarray}
Likewise, if the variance $\overline{\delta \varepsilon_K^2} = \overline{\varepsilon^2_K - \bar{\varepsilon}_K^2}$ is 
given, we also require 
 \begin{eqnarray}
\overline{\delta \varepsilon_K^2} = \frac{(k_{\rm B}T)^2}{2}   \Rightarrow T = \sqrt{\frac{2}{k_{\rm B}}} \sqrt{\overline{\delta\varepsilon_K^2}}
\label{eq:variance}
\end{eqnarray}

Thus given a set of kinetic energies of a particle, we can estimate the parameter $T$ using either the mean value (Eq.~\ref{eq:mean}) or the variance (Eq.~\ref{eq:variance}). Note that the variance is but a measure of the heat capacity $C_v$ (of the ideal gas). Consistency between these two estimates, the two moments of Eq.~\ref{eq:MB}, is a good indicator of adequate sampling. 

In comparing observed data of kinetic energies from a simulation versus Eq.~\ref{eq:MB}, one could construct the cumulative distribution from the observed data and the cumulative distribution from Eq.~\ref{eq:MB} for various choices of $T$ and ask for the value of $T$ that minimizes the least square deviation between the simulated and predicted distribution. In exploratory calculations, we found comparing $T$ from the two paths noted above works as well as obtaining a $T$ to fit the cumulative distribution function. Thus we estimate $T$ from Eqs.~\ref{eq:mean} and ~\ref{eq:variance}.

 \subsection{Translational and rotational kinetic energies}

Assume we have the site velocity $\mathbf{v}_i$ for each site $i$ (=O, H$_1$, H$_2$) of a water molecule. From the site velocities, we can construct the velocity of the center of mass, $\mathbf{v}_{com}$. Then the translational kinetic energy of the 
water molecule is simply $(M/2) \mathbf{v}_{com} \cdot \mathbf{v}_{com}$, where $M$ is the mass of the water molecule. 
	
For calculating the rotational kinetic energy we need the angular velocity, $\boldsymbol{\omega}$. To simplify the notation, without loss of generality we assume $\mathbf{v}_i$ is relative to the velocity of the center of mass. For a chosen water molecule we have, for each site $i$, 
	\begin{equation}
		\boldsymbol{\omega} \times \mathbf{r}_i = \mathbf{v}_i.
	\end{equation}
	But a direct application of the above equation cannot be used to calculate $\boldsymbol{\omega}$ because the matrix equation is singular. We therefore define\cite{singer:jcp2018b,palmer:jcp2018}
	\begin{equation}
		\mathcal{L} = \sum_i | \mathbf{v}_i - \boldsymbol{\omega} \times \mathbf{r}_i |^2.
	\end{equation}
	We minimize $\mathcal{L}$ with respect to the $x, y, z$-components of $\boldsymbol{\omega}$, giving the following matrix equation:
 
	\begin{equation}
		\begin{pmatrix}
			\sum_{i} z_i^2 + y_i^2  & -\sum_i x_i y_i           & -\sum x_i z_i \\
			-\sum_i x_i y_i            & \sum_i x_i^2 + z_i^2  & -\sum y_i z_i \\
			-\sum_i x_i z_i            & -\sum y_i z_i               & \sum_i x_i^2 + y_i^2
		\end{pmatrix}
		\begin{pmatrix}
			\omega_x \\
			\omega_y \\
			\omega_z 
		\end{pmatrix}
		 = 
		\begin{pmatrix}
			\sum_i (\boldsymbol{r}_i \times \boldsymbol{v}_i)_x \\
			\sum_i (\boldsymbol{r}_i \times \boldsymbol{v}_i)_y \\
			\sum_i (\boldsymbol{r}_i \times \boldsymbol{v}_i)_z 
		\end{pmatrix}
	\end{equation}
	In tensor notation, we can write this more compactly as
	\begin{equation}
		r_j \omega_p r_q \epsilon^{ljk} \epsilon_{pqk} = r_j v_k \epsilon_{jkl},
	\end{equation}
	summed over all sites. 
	
	We constructed the above matrix for each molecule and solved for $\boldsymbol{\omega}^\prime = \{\omega_x, \omega_y, \omega_z\}$. Then the rotational kinetic energy is $0.5 \boldsymbol{w}^\prime\cdot \overline{\overline{I}}\cdot \boldsymbol{w}$, where $\overline{\overline{I}}$ is the moment of inertia tensor and the transpose of the vector is indicated by a prime.

\section{Methodology}

\subsection{Simulations}

The system comprises  $N=2006$ SPC/E\cite{spce} water molecules. The volume was adjusted to match the experimental density\cite{nist} --- 33.33 water molecules/nm$^3$ or 0.997 g/cm$^3$ --- of liquid water at 298.15~K and 1 atm pressure. The system was simulated at constant volume at a thermostat temperature of 298.15~K. We used two simulation packages, NAMD \cite{namd,namd:2020} and LAMMPS\cite{plimpton:jcop1995,lammps:2022}, for performing the simulations.  (We note in passing that results based on exploratory calculations with OpenMM \cite{openmm} are also consistent with the main findings based in this paper.) The bulk of our work is based on simulations using NAMD with the 2006 water system.  

To check system size effects, using the LAMMPS code we conducted a limited set of simulations starting from a previously well-equilibrated box of 16384 water molecules. Results obtained using this system are noted as such.

\subsubsection{NAMD}
The force due to Lennard-Jones interactions between oxygen atoms was smoothly switched to zero between 9 {\AA} and 10 {\AA}. Electrostatic interactions were described using the particle mesh Ewald (PME) method. The 
grid spacing for PME was 1.0 {\AA} and the {\sc PMETolerance} is set to $10^{-6}$.  (These are the recommended defaults within NAMD.) 
Within NAMD, we use the default choice of the {\sc SETTLE} \cite{kollman:settle92} algorithm to constrain the geometry of a water molecule. 
Molecular dynamics simulations at constant temperature were performed using two different thermostats, the stochastic velocity rescaling (SVR) \cite{svr:jcp07} and Langevin thermostats, respectively. The time period for stochastic velocity rescaling was set to 1~ps. (The NAMD manual recommends values between 0.5 to 2~ps, with larger values leading to more NVE-like dynamics.) For Langevin dynamics, the Langevin damping was set to 1 ps$^{-1}$. The  friction due to the heat bath is applied only to the oxygen atoms, i.e.\ {\sc langevinhydrogen off} is used.  For NVE simulations, everything is maintained as discussed above except that the thermostat is removed. 

Dr.\ Valiya Parambathu (University of Delaware) conducted exploratory calculations with OpenMM\cite{openmm} using default Langevin thermostat settings and shared the data with us. This code, like NAMD, also uses SETTLE. The overall conclusions in this work remain the same. 

\subsubsection{LAMMPS}
We also perform a limited set of runs using LAMMPS to validate the central observations using NAMD.  For LAMMPS, to be consistent with the NAMD run, we use CHARMM style force-switching {\sc charmmfsw} with the LJ forces being switched to zero between 9 {\AA} and 10 {\AA}. For electrostatics, we set {\sc kspace\_style pppm 5e-6}, i.e.\ we require the error relative to the Coulomb force to be $5\times 10^{-6}$. (Please note that {\sc PMETolerance} in NAMD is based on the Coulomb energy at the cutoff.) We use the SVR thermostat with a period of 1~ps (1000~fs). We also conduct a limited set of runs with  {\sc kspace\_style pppm} set to $10^{-6}$ and $10^{-7}$. The central results remain the same. 

\textbf{Rigid body dynamics vs.\ constrained dynamics}: Within LAMMPS we also simulate a system with 16384 water molecules using two different descriptions: treating the system as a collection of rigid objects and using rigid body integration or, as with NAMD, treating the system as a collection of molecules each of whose geometry is constrained by RATTLE\cite{andersen:rattle83} to maintain rigidity. This test ensures that the formulation of the equations of motion are not the reason for the failure of equipartition noted in the main text. (To appreciate the subtlety implied here, note that for an atom that is constrained to rotate in the $x-y$ plane about a fixed point at a constant angular velocity of $\omega$, 
the kinetic energy versus time obtained by integrating the velocity Verlet equation for the atom will be oscillatory, as is expected of the integrator. However, the rotational kinetic energy using the same site velocity will be constant. The deviation between the two measures of the kinetic energy is proportional to $\delta t^4$.) As expected, both the formulations predict the failure of equipartition noted in the main text.

\subsubsection{Time steps and simulation log}
 
 The original $N=2006$ water simulation cell is taken from a well thermalized system from an earlier study.  For a given time-step $\delta t$ for integrating the equations of motion, we further equilibrate the system for $4\times 10^6$ time steps and then collect data for $10 \times 10^6$ time steps. The production run is broken into two steps of $5\times 10^6$ steps each with the second step started using restart files. This is done for convenience in submitting jobs on a single node of the Summit cluster.
In the production phase, we save configurations and velocities every 250 time steps for further analysis. This gives us $40\times 10^3$ frames to analyze. We study $\delta t = 0.5$~fs to $\delta t = 3.0$~fs in steps of 0.5~fs. 

With NAMD, we output both the coordinates and the velocities as DCD files. We used the 
MDTRAJ\cite{mdtraj}  package to concatenate the DCD files between the two steps noted above. 
With LAMMPS, we output the coordinate file as a DCD but save the velocities as an ASCII file using the {\sc dump} option.
LAMMPS also allows dumping the angular velocities of a defined collection of atoms. In exploratory calculations, we use this option to cross-check our calculation of the angular velocity (see below): the agreement between our calculation and that of LAMMPS is excellent. Note that in our earlier study on spin-rotation relaxation in CH$_4$ we followed a similar procedure \cite{singer:jcp2018b}; indeed, the code for angular velocities used in this work is derived from that earlier work. 
 
 For the NVE runs with NAMD, we take the last configuration of the $\delta t = 0.5$~fs run with the SVR thermostat and launch constant energy runs for 20~ps saving velocities and positions every time step. We use the same seed for the random number generator for the different $\delta t$ cases.
 
\subsubsection{Potential energy calculation}\label{sc:be}

We use the {\sc PAIRINTERACTION} facility within NAMD to compute the interaction energy between a distinguished water molecule and the rest of the fluid. We sample a total of 100 water molecules from the system. In the past, for example Ref.~\citenum{weber:jcp11}, we used our in-house Python codes to write the PDB files to indicate the two groups of atoms between which the pair interaction is needed. In the present study, we use the MDAnalysis package for convenience\cite{mdanalysis:2011, mdanalysis:2016}. 

For water molecule $i$ we have $40\times 10^3$ binding energy values.  We find the mean $\bar{\varepsilon}_i$ of the binding energy, and from this, for the sample of 100 water molecules, we compute the overall mean $\langle\varepsilon\rangle$ and the standard error of the mean. This is the data shown in Figure~4(a) of the main text. 

If our sample size is adequate, $N \langle \varepsilon\rangle / 2$ should be close to the mean potential energy of the entire system. The simulation log-file already provides the potential energy at  a defined time point. From the time-ordered potential energy values, we can compute the mean and standard error; we use the Friedberg and Cameron \cite{friedberg:1970} approach to take care of any correlation in the time-ordered data. Another way to find the correlation length in the time-ordered data is to compute an autocorrelation function and treat values above a defined cutoff (we use 0.05) as correlated. These approaches yield very similar results and for convenience we use the Friedberg-Cameron approach. 

Since we have only 40,000 overall potential energy values, we extend the NVT simulations another $10 \times 10^6$ steps, and compute the mean and uncertainty of the $80,000$ potential energy values for the entire system.  Within statistical uncertainties the estimate of the potential energy of the simulation cell $N \langle \varepsilon\rangle / 2$ agrees with the mean value based on the simulation log file. 

\subsection{Calculation of single molecule, rigid body kinetic energies}\label{sc:rottrans}

For obtaining $T$, we need to sample molecules and obtain their translational and rotational kinetic energies. To this end, we sample 1050 molecules out of the $2006$ water molecules, i.e.\ roughly half the number of water molecules in the simulation system. We use MDTRAJ\cite{mdtraj}  to extract from the corresponding DCD file the coordinates and velocities of a chosen water molecule and write it as a separate DCD file. (Being able to extract one water molecule at a time allows us to parallelize the subsequent calculation over all the cores that are available.)  

Using the coordinates and velocities, and the ideas in Section 1, we obtain the translational and rotational kinetic energy of the chosen water molecule for all the $40\times 10^3$ frames. (When obtaining velocities from the NAMD velocity DCD file, we take care to multiply the values by the {\sc PDBVELFACTOR} noted in the NAMD user guide to get velocities in units of {\AA}/ps.) 
From this data, we estimate $T$ using either Eq.~\ref{eq:mean} or Eq.~\ref{eq:variance}. For the collection of 1050 molecules, we then obtain the mean $T$ and the standard error of the mean for either approach. 

For LAMMPS, we found it most convenient to reformat the ASCII velocity data to a DCD (using a combination of MDAnalysis for writing a DCD and MDTRAJ for concatenating DCDs). We take care to ensure units and multiplicative factors are correctly included. The analysis then proceeds as noted above. 

We check our calculations two different ways. First, the kinetic energy obtained using the velocity of each atom in the water molecule should equal the sum of the translational and rotational kinetic energies. This check is validated. Extending this check to all the water molecules in the system, the kinetic energy printed in the simulation log file should equal that based on the sum of the translational and rotational kinetic energies for all the water molecules. This check is also validated. In particular, with LAMMPS, these consistency checks agree to the number of decimals used in the LAMMPS log file, while for NAMD the agreement is usually good to at least 3 decimal places.  (Part of the excellent agreement in the case of LAMMPS may be due to the use of RATTLE versus SETTLE. See Sec.\ \ref{sc:settle} below.) 

For LAMMPS runs with the larger $N=16384$ system, we sampled 8190 water molecules, i.e.\ roughly half the number of water molecules in the system. The rest of the procedure is as above.

\subsection{Hydration free energy of a soft, repulsive cavity}\label{sc:qct}

We create a soft, repulsive cavity about the center of the simulation cell by applying the WCA potential of the form \cite{Weber:jctc12}
\begin{eqnarray}
\phi (r; \lambda) & = & 4a\bigg[ \left(\frac{b}{r-\lambda+\sqrt[6]{2}b}\right)^{12} - \left(\frac{b}{r-\lambda+\sqrt[6]{2}b}\right)^{6} \bigg] + a \label{eq:lj} \; ,
\end{eqnarray}
where $a = 0.155$~kcal/mol and $b = 3.1655$~{\AA} are positive constants and  ($r < \lambda$). For  $r \geq \lambda$, $\phi (r) = 0$.  As in our earlier studies, we use the Tcl-interface to NAMD \cite{namd} to impose the force due to $\phi$
on the oxygen atoms of the water molecule. 

The free energy, $\mu^{{\rm ex}}$, to create a cavity of radius $\lambda_f$ is given by 
\begin{eqnarray}
\mu^{{\rm ex}} = \int_0^{\lambda_f} \left\langle \frac{d \phi(r; \lambda)}{d\lambda} \right\rangle_\lambda \, d\lambda 
\end{eqnarray}
Here $\lambda_f = 4$~{\AA}. We estimate the integral using Gauss-Legendre quadratures  \cite{Hummer:jcp96}.  In earlier studies we have used a 7-point quadrature for every unit {\AA}ngstr{\"o}m. Here we split the domain two different ways: 
(1) over $[0,2]$~{\AA}, $[2,3]$~{\AA}, and $[3,4]$~{\AA}, and for each interval used a 5-point Gauss-Legendre quadrature, and (2) over $[0,2]$~{\AA} using 5 quadrature points and $[2,4]$~{\AA} using 9 quadrature points. At each 
quadrature point, we simulate the system for 1.8~ns saving the force values every 50~fs. Data from the last 1~ns is 
used in the analysis.  Error analysis and error propagation is performed as before \cite{Weber:jctc12}: The standard error of the mean force is obtained using the Friedberg-Cameron algorithm \cite{friedberg:1970,allen:error} and in adding multiple quantities, the errors are propagated using standard variance-addition rules. Finally, the free energy estimates from either partitioning is averaged and the errors propagated appropriately. 

We calculate the free energy for temperatures 282.15~K, 290.15~K, 298.15~K, 306.15~K, and 314.15~K. The volume of the simulation cell at each temperature is based on the experimental density\cite{nist} of the liquid at that temperature. The $\mu^{{\rm ex}}$ values thus obtained are noted in Fig.~4(b) main text. 

\newpage

\section{Supplemental Results}

\subsection{Kinetic energies --- NAMD \& LAMMPS}
\setlength{\tabcolsep}{12pt}
\newcommand{\TT}[2]{\ensuremath{{#1}\pm{#2}}}
\ctable[
	mincapwidth=\textwidth,
	caption={Translational and rotational kinetic energies (expressed in degree Kelvin) for the stochastic velocity rescaling (SVR) thermostat. 
	The temperature obtained using $\bar{\varepsilon}_K$, the mean kinetic energy, is $T(\bar{\varepsilon}_K)$, and that obtained using $C_v$, the ideal gas heat capacity, is $T(C_v)$. Data here is graphed in Fig.~1 (main text).},
	label=tb:namdsvr,
	pos=h,
	captionskip=-1.ex
	]	
{c c c c c}
{}
{
\FL
	                  &                   \multicolumn{2}{c}{Translation}              &                   \multicolumn{2}{c}{Rotation}                      \NN[-1ex]
	                                            \cmidrule(rl){2-3}                                                               \cmidrule(rl){4-5}                       
   $\delta t$ (fs)  &  $T(\bar{\varepsilon}_K)$        &        $T(C_v)$        &    $T(\bar{\varepsilon}_K)$        &      $T(C_v)$           \ML
0.5                     & \TT{297.9}{0.04}                    & \TT{297.9}{0.06}   &    \TT{298.0}{0.04}                   &    \TT{298.0}{0.06} \NN[-1ex]
1.0                     & \TT{298.3}{0.04}                    & \TT{298.3}{0.06}   &    \TT{297.9}{0.04}                   &    \TT{297.9}{0.06} \NN[-1ex]
1.5                    & \TT{298.6}{0.04}                    & \TT{298.5}{0.06}   &    \TT{297.4}{0.04}                   &    \TT{297.3}{0.06} \NN[-1ex]
2.0                    & \TT{299.2}{0.04}                    & \TT{299.2}{0.06}   &    \TT{297.1}{0.04}                   &    \TT{296.9}{0.06} \NN[-1ex]
2.5                    & \TT{299.9}{0.04}                    & \TT{300.0}{0.06}   &    \TT{296.3}{0.04}                   &    \TT{296.2}{0.06} \NN[-1ex]
3.0                    & \TT{300.6}{0.04}                    & \TT{300.6}{0.06}   &    \TT{295.3}{0.04}                   &    \TT{295.1}{0.06} \LL
}

\ctable[
	mincapwidth=\textwidth,
	caption={Translational and rotational kinetic energies (expressed in degree Kelvin) for the Langevin thermostat. Rest as in Table~\ref{tb:namdsvr}.},
	label=tb:namdlangevin,
	pos=h,
	captionskip=-1.ex
	]	
{c c c c c}
{}
{
\FL
	                  &                   \multicolumn{2}{c}{Translation}              &                   \multicolumn{2}{c}{Rotation}                      \NN[-1ex]
	                                            \cmidrule(rl){2-3}                                                               \cmidrule(rl){4-5}                       
   $\delta t$ (fs)  &  $T(\bar{\varepsilon}_K)$        &        $T(C_v)$        &    $T(\bar{\varepsilon}_K)$        &      $T(C_v)$           \ML
0.5                     & \TT{298.2}{0.04}                    & \TT{298.2}{0.06}   &    \TT{298.1}{0.04}                   &    \TT{298.0}{0.06} \NN[-1ex]
1.0                     & \TT{298.1}{0.04}                    & \TT{298.1}{0.06}   &    \TT{297.7}{0.04}                   &    \TT{297.7}{0.06} \NN[-1ex]
1.5                    & \TT{298.1}{0.04}                    & \TT{298.2}{0.06}   &    \TT{296.9}{0.04}                   &    \TT{296.8}{0.06} \NN[-1ex]
2.0                    & \TT{298.3}{0.04}                    & \TT{298.3}{0.06}   &    \TT{296.1}{0.04}                   &    \TT{296.0}{0.06} \NN[-1ex]
2.5                    & \TT{298.3}{0.04}                    & \TT{298.3}{0.06}   &    \TT{294.8}{0.04}                   &    \TT{294.6}{0.06} \NN[-1ex]
3.0                    & \TT{298.4}{0.04}                    & \TT{298.4}{0.06}   &    \TT{293.4}{0.04}                   &    \TT{293.1}{0.06} \LL
}

\ctable[
	mincapwidth=\textwidth,
	caption={Translational and rotational kinetic energies (expressed in degree Kelvin) for the SVR thermostat and the LAMMPS code. 
	Rest as in Table~\ref{tb:namdsvr}.},
	label=tb:lammpssvr,
	pos=h,
	captionskip=-1.ex
	]	
{c c c c c}
{}
{
\FL
	                  &                   \multicolumn{2}{c}{Translation}              &                   \multicolumn{2}{c}{Rotation}                      \NN[-1ex]
	                                            \cmidrule(rl){2-3}                                                               \cmidrule(rl){4-5}                       
   $\delta t$ (fs)  &  $T(\bar{\varepsilon}_K)$        &        $T(C_v)$        &    $T(\bar{\varepsilon}_K)$        &      $T(C_v)$           \ML
1.0                     & \TT{298.3}{0.04}                    & \TT{298.3}{0.06}   &    \TT{297.8}{0.04}                   &    \TT{297.8}{0.06} \NN[-1ex]
2.0                    & \TT{299.2}{0.04}                    & \TT{299.2}{0.06}   &    \TT{297.0}{0.04}                   &    \TT{296.9}{0.06} \LL
}

\subsection{Relative velocity along constrained bonds}\label{sc:settle} 

As we were doing the calculations noted in Sec.~\ref{sc:rottrans}, we also decided to check the impact of the constraint algorithms --- SETTLE\cite{kollman:settle92} or RATTLE\cite{andersen:rattle83}. To this end, let $\mathbf{v}_O$ and $\mathbf{v}_H$ be the velocities of the oxygen atom and one of the hydrogen atoms, respectively, in a chosen water molecule. Let $\mathbf{r}_O$ and $\mathbf{r}_H$ be the respective coordinates. Then, since the O-H bond is constrained, we require
\begin{eqnarray}
	\Delta_{v} = (\mathbf{v}_H - \mathbf{v}_O)\cdot \mathbf{\hat{r}}_{OH} = 0 \, ,
\end{eqnarray}
where $\mathbf{\hat{r}}_{OH} = (\mathbf{r}_H - \mathbf{r}_O) / || (\mathbf{r}_H - \mathbf{r}_O) || $ is the unit vector along the O-H bond. We can then study the distribution of $\Delta_v$ for the sampled water molecules. 

\begin{figure}
	\includegraphics[scale=1]{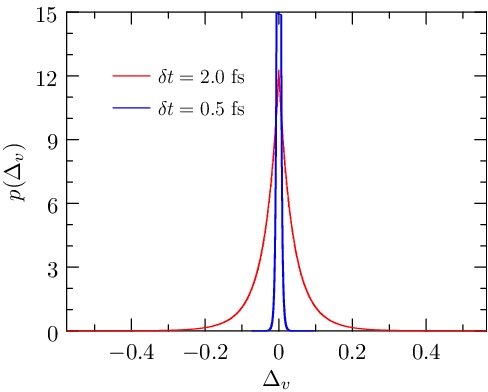}
\caption{Distribution of $\Delta_v$ using the SETTLE procedure in NAMD for two different time steps. On the same scale, $p(\Delta_v)$ obtained using RATTLE in LAMMPS is close to a delta function.}\label{fg:bond}
\end{figure}

Fig.~\ref{fg:bond} shows that the SETTLE procedure can introduce errors in that the above condition is 
is not strictly met. 
However, the impact of this discrepancy is not significant for either the conservation of energy (in the numerical integration scheme), which is where one would suspect it would most impact, or in the primary conclusions in the main text.

\subsection{Sampling of the shadow Hamiltonian} 
For a system with the physical Hamiltonian $\mathcal{H}$, the discretized numerical solution follows a so-called shadow Hamiltonian\cite{toxvaerd:pre94,gans:pre2000,shaw:jctc2010} $\tilde{\mathcal{H}}$ that is related to $\mathcal{H}$ by 
\begin{eqnarray}
	\tilde{\mathcal{H}}(\delta t) = \mathcal{H} + \mathcal{O}(\delta t^2)
\end{eqnarray}	
This structure confounds the relation between momentum and velocity in classical mechanics\cite{gans:pre2000}; however, since both NAMD and LAMMPS use the on-step velocity (from the velocity Verlet integration) procedure for calculating temperatures, we will ignore this complication here.  Accepting this, the standard deviation in the energy, $\sigma(\delta t)$, in a formal NVE simulation with time-step $\delta t$ is expected to scale as \cite{shirts:plos18} $\sigma(\delta t) \sim \delta t^2$ and thus 
\begin{eqnarray}
	\frac{\sigma(\delta t_1)}{\sigma(\delta t_2)} = \frac{\delta t_1^2}{\delta t_2^2}
\end{eqnarray}
Relative to $\delta t_2 = 1$~fs, we then have 
\begin{eqnarray}
	\sigma_{rel}(\delta t) = \frac{\sigma(\delta t)}{\sigma(\delta t_2=1)} = \delta t^2
	\label{eq:scaling}
\end{eqnarray}

\begin{figure}
	\includegraphics[scale=1.0]{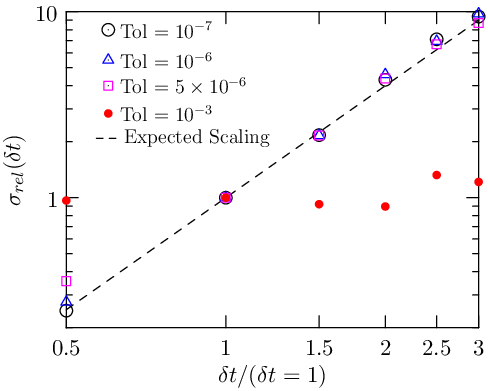}
\caption{Scaling of the standard deviation of the simulated energy in an NVE run using LAMMPS and a 16384 water molecule system. 
{\sc Tol} specifies the relative tolerance of forces specified in the PPPM method. Tolerance values between $10^{-5}$ and $10^{-6}$ are usual; most of our production runs with LAMMPS used either $5\times 10^{-6}$ or $10^{-6}$. The results using a very poor tolerance of $10^{-3}$ are shown only for illustrating the sensitivity of Eq.~\ref{eq:scaling} (the expected scaling behavior).}\label{fg:lmp_fluc}
\end{figure}

Figure~\ref{fg:lmp_fluc} shows the behavior of this scaling relation obtained using the 16384 system simulated within LAMMPS. The standard deviation of energy is obtained from a short 500 step NVE simulation that followed at 1000 step NVT simulation. (Energies are saved to the log file at each time step.) 

\begin{figure}
\includegraphics[scale=0.95]{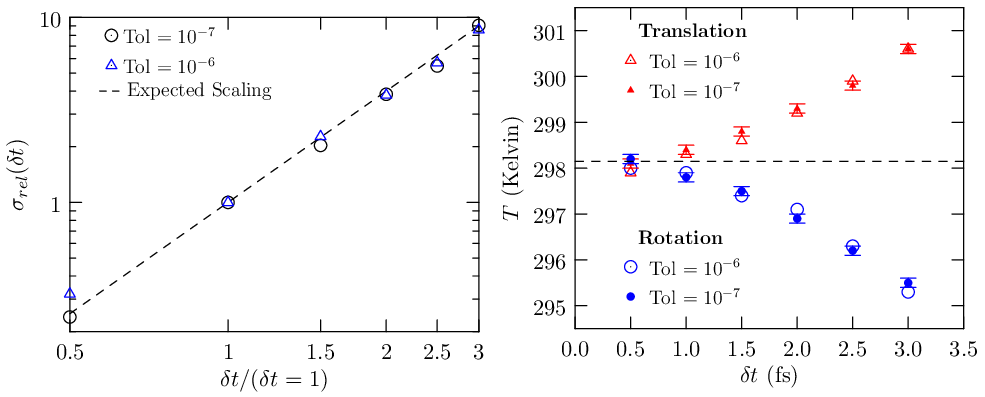}
\caption{\underline{Left panel}: Scaling of the standard deviation of the simulated energy in an NVE run using NAMD and a 2006 water molecule system. The system is thermalized using the SVR thermostat for $2\times 10^6$ steps and the NVE run conducted for 500 steps with energy logged every step. For particle mesh Ewald simulations, the recommended (default) tolerance in NAMD, {\sc PMEtolerance}, is $10^{-6}$ and the (default) {\sc PMEgridspacing} is 1.0~{\AA}.  We also test a very tight tolerance of $10^{-7}$ and a 0.5~{\AA} grid spacing. \underline{Right panel}: The dependence of translation and rotation temperature for a $2\times 10^6$ production run following the aforementioned equilibration run. The SVR thermostat is used throughout. The results with the default Ewald tolerance (Fig.~1, main text) obtained for a long simulation are consistent with the results obtained using the tight tolerance.}\label{fg:namd_fluc}
\end{figure}

Figure~\ref{fg:namd_fluc} shows the behavior of the scaling relation (Eq.~\ref{eq:scaling}) for the 2006 water molecule system simulated using NAMD.  We also compare the behavior of the translation and rotation temperatures obtained noted in Fig.~1 (main text) with results obtained using a shorter run but with a very tight tolerance for the particle mesh Ewald calculation. 

\clearpage
\newpage
\subsection{Temperature distribution in NVE simulations with LAMMPS }

\begin{figure}
\includegraphics[scale=0.95]{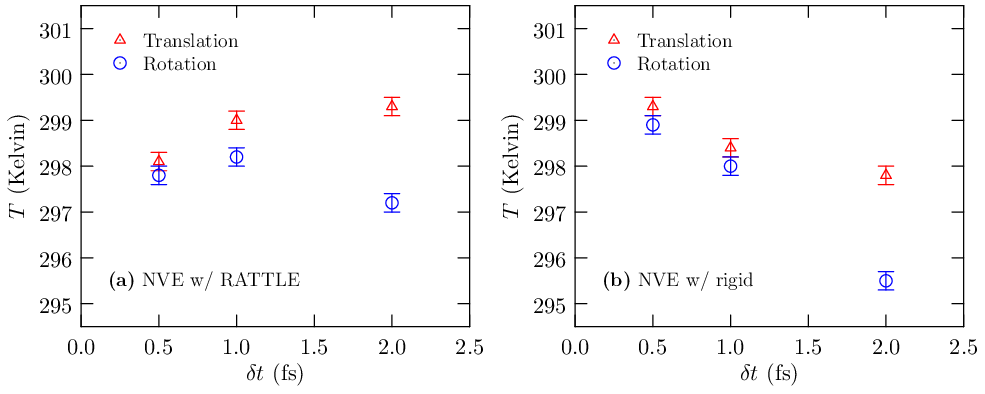}
\caption{Behavior of translational and rotational temperature in an NVE simulation, i.e.\ with no thermostat applied. The $2\sigma$ standard error of the mean is shown. \underline{Left panel}: Water molecules are constrained using RATTLE.  \underline{Right panel}: Water molecules are described as rigid bodies.}\label{fg:lmp_nve}
\end{figure}

Figure~\ref{fg:lmp_nve} shows the behavior of the translational and rotational temperature in an NVE simulation, i.e.\ without any thermostat. The system comprises 16384 water molecules of which 8190 were sampled. For Fig.~\ref{fg:lmp_nve} (left panel), the system is first equilibrated over 50,000 steps using the {\sc nvt} option. The run is continued for another 100,000 steps without any thermostat. A production run of 350,000 steps followed with data being saved every 250 steps. For Fig.~\ref{fg:lmp_nve} (right panel), there are four stages: (1) 50,000 steps of equilibration under {\sc nvt} with water geometry constrained using RATTLE; (2) 50,000 steps of equilibration under {\sc rigid/nvt/small}, i.e.\ describing the collection of water molecules as rigid objects; (3) 50,000 steps of equilibration under {\sc rigid/nve/small}, i.e.\ describing the collection of water molecules as rigid objects but without any thermostat; and (4) 350,000 steps of production under {\sc rigid/nve/small} with data saved every 250 steps for analysis.  We use {\sc kspace\_style pppm 5e-6} for all these simulations. 
We note that NVE runs with NAMD (and the 2006 water system) also display similar behavior. 

\begin{figure}
	\includegraphics[scale=0.95]{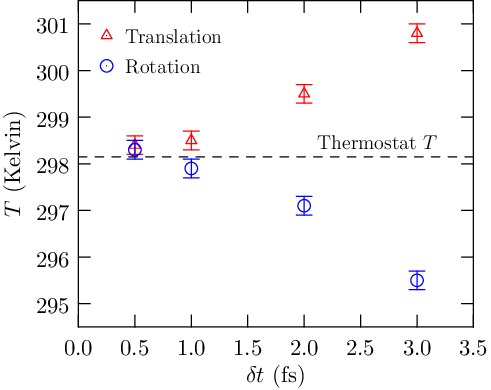} 
\caption{Behavior of translational and rotational temperature in a 16384 water simulation with the SVR thermostat. Water molecules are constrained using RATTLE. The $2\sigma$ standard error of the mean is shown.}\label{fg:lmp_svr}
\end{figure}

As a further check, for the same 16384 water molecule system noted above, we also examine the translational and rotational temperature in the presence of the SVR thermostat. For these runs, the system is equilibrated for 250,000 steps with {\sc RATTLE}, followed by a production phase of 500,000 steps with data saved every 250 steps. The results shown in Fig.~\ref{fg:lmp_svr} recapitulates the trends seen within NAMD with 2006 water molecules or LAMMPS with 2006 water molecules. 

The above results emphasize that the properties of the thermostat (or our choice of thermostat parameters) or the formulation of the equations of motion are not the reason for the failure of equipartition noted in the main text.  

\subsection{Influence of proton mass for $\delta t = 4$~fs}
There have been efforts to use $\delta t = 4$~fs by changing the mass of the hydrogen atom in the rigid water models. The larger $\delta t$ of course facilitates longer total simulation times. Our results (main text) however suggest that this approach may suffer from the failure of equipartition. To test this, we simulate the 16384 water molecule system with $\delta t = 4$~fs. Proton masse of 2, 3, and 4 (a hypothetical atom) are tested. For each chosen proton mass, we change the mass of the oxygen atom to ensure that the net mass of the water molecule remains the same as for liquid water (18.0154 amu). 

\begin{figure}
	\includegraphics[scale=0.95]{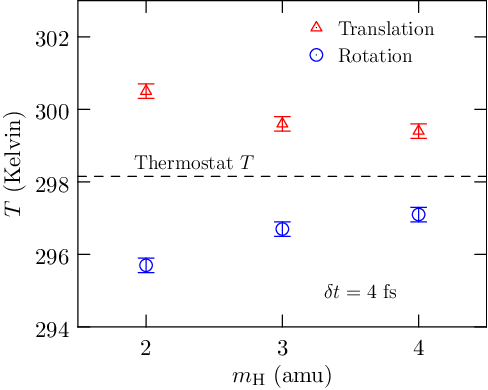}
\caption{Behavior of translation and rotation temperature for a simulation with 16384 water molecules with the SVR thermostat using LAMMPS.  Of the 16384 water molecules, 8190 water molecules were sampled for estimating the temperature using the procedure in Sec.~2.2.}\label{fg:deshaw}
\end{figure}

The system is equilibrated for $250 \times 10^3$ steps and data collected over $500 \times 10^3$ time steps. Frames are saved every 250 time steps for further analysis. The procedure for calculating the translation and rotation temperature is as noted in Sec.~2.2. Figure~\ref{fg:deshaw} shows that changing the hydrogen mass can help alleviate the failure of equipartition. However, even with a hydrogen mass of 4 amu, the failure of equipartition persists. Thus our results suggest caution in using an aggressively large $\delta t$ in an effort to span longer overall simulation times. 

\subsection{Mean-squared displacement} 
\begin{figure}
	\includegraphics[scale=1.0]{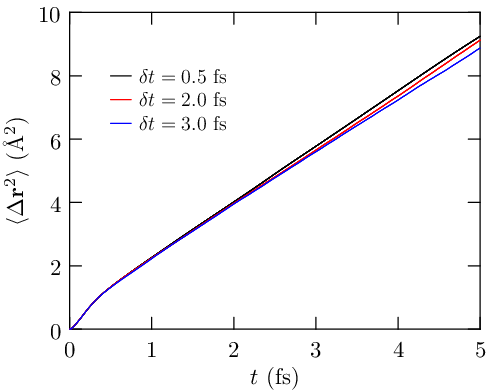}
\caption{Mean-squared displacement versus time for select $\delta t$ values. The computed diffusivities are $0.29$ {\AA}$^2$/ps, $0.29$ {\AA}$^2$/ps, and $0.27$ {\AA}$^2$/ps, respectively, for $\delta t = 0.5, 2.0, \mathrm{and}\; 3.0$~fs.}\label{fg:msd}
\end{figure}

Figure~\ref{fg:msd} shows the mean-squared displacement for select $\delta t$ values. For the reference $\delta t = 0.5$~fs data, the integral of the velocity autocorrelation (Fig.~2, main text) out to 0.8 ps converges to the diffusion coefficient 
obtained using the mean-squared displacement.  
 Note also the obvious oddity that the deviation in the initial velocity ACFs (Fig.~3, main text) 
 for translational motion, relative to the $\delta t = 0.5$~fs reference, are all positive, whereas the diffusion coefficient assessed from the mean squared displacements for $\delta t > 0.5$~fs are all smaller than the value obtained using $\delta t = 0.5$~fs. 
This just emphasizes that discretization errors can get buried when we compute integral quantities such as the diffusion coefficient from the velocity ACF or from the slope of the mean-squared displacements. 

\clearpage
\newpage

\providecommand{\latin}[1]{#1}
\makeatletter
\providecommand{\doi}
  {\begingroup\let\do\@makeother\dospecials
  \catcode`\{=1 \catcode`\}=2 \doi@aux}
\providecommand{\doi@aux}[1]{\endgroup\texttt{#1}}
\makeatother
\providecommand*\mcitethebibliography{\thebibliography}
\csname @ifundefined\endcsname{endmcitethebibliography}
  {\let\endmcitethebibliography\endthebibliography}{}